# Graphdiyne as a Promising Substrate for Stabilizing Pt Nanoparticle Catalyst

Zheng-Zhe Lin*

*School of Physics and Optoelectronic Engineering, Xidian University, Xi'an 710071, China*

*Corresponding Author. E-mail address: linzhengzhe@hotmail.com

**Key Words:** graphdiyne, Pt nanoparticle

PACS **61.46.-w**   Structure of nanoscale materials

PACS **68.65.-k**   Low-dimensional, mesoscopic, nanoscale and other related systems: structure and nonelectronic properties

**Abstract** – At present, Pt nanoparticle catalysts in fuel cells suffer from aggregation and loss of chemical activity. In this work, graphdiyne, which has natural porous structure, was proposed as substrate with high adsorption ability to stabilize Pt nanoparticles. Using multiscale calculations by *ab initio* method and the ReaxFF potential, geometry optimizations, molecular dynamics simulations, Metropolis Monte Carlo simulations and minimum energy paths calculations were performed to investigate the adsorption energy and the rates of desorption and migration of Pt nanoparticles on graphdiyne and graphene. According to the comparison between graphdiyne and graphene, it was found that the high adsorption ability of graphdiyne can avoid Pt nanoparticle migration and aggregation on substrate. Then, simulations indicated the potential catalytic ability of graphdiyne-Pt-nanoparticle system to the oxygen reduction reaction in fuel cells. In summary, graphdiyne should be an excellent material to replace graphite or amorphous carbon matrix for stabilizing Pt nanoparticle catalysts.

## I. Introduction

In modern society, the demand for clean, efficient energy source is increasingly urgent. At present, in the widely used internal-combustion engines and aero-engines chemical energy is converted into heat and then into mechanical energy. But the efficiency of the latter process is limited by the second law of thermodynamics. To solve this problem, fuel cells, which directly convert chemical energy into electric energy, are considered as next-generation energy source for powering stationary systems, portable electronic devices, and vehicles. Furthermore, fuel cells with hydrogen as a fuel are clean energy sources without causing environment contamination. In fuel cells, Pt nanoparticles with a size of 2~5 nm adsorbed on high surface area amorphous carbon matrix are proved to be excellent catalyst. However, the Pt nanoparticles suffer from migrating and coarsening [1], resulting in aggregation and loss of surface area and chemical activity. The instability of Pt nanoparticles has become a major limitation for commercialization of fuel cell technology. Several solutions have been considered for this problem so far, e.g. using stable Pt-bimetallic [2] or multimetallic [3] catalysts. But these solutions are still far beyond industrialization.

To solve the problem of Pt nanoparticle aggregation in the catalysts, it is necessary to improve the adsorption capability of the substrate. Changing the structure of carbon matrix may be a convenient solution. For example, carbon nanowalls are proposed to be separate Pt supports for fuel cells [4], and the adsorption and catalytic ability of Pt nanoparticles were found enhanced on defective graphene [5]. Recently, a new carbon allotrope named graphdiyne, which was theoretically predicted more than 20 years ago [6, 7], has been successfully synthesized on the surface of copper via a cross-coupling reaction using hexaethynylbenzene [8]. Graphdiyne may be the most stable of all artificially synthesized carbon allotropes [9], and has shown an improved performance in polymer solar cells [10]. Since graphdiyne has a porous structure (Fig. 3(a)), Pt nanoparticles may be embedded with a stronger adsorption energy than that on graphite/graphene. In addition, the high electronic conductivity and mobility of graphdiyne [11, 12] makes it suitable to be electrodes of fuel cells (e.g. see Ref. [13]). And the unique mechanical property of graphdiyne against armchair and zigzag loads suggest flexible designations towards functional use of this novel material [14].

In this work, the adsorption potential of graphdiyne to Pt nanoparticles was theoretically studied and compared with the adsorption potential of graphene. Using multiscale calculations by *ab initio* method and the ReaxFF potential [15-18], geometry optimizations, molecular dynamics (MD) simulations, Metropolis Monte Carlo (MMC) simulations and minimum energy path (MEP) calculations were performed to investigate the adsorption energy, the desorption and migration rate of Pt nanoparticles on graphdiyne and graphene. According to the result, the adsorption of Pt nanoparticles on graphdiyne is much stronger than on graphene. On graphdiyne surface, Pt nanoparticles are tightly bound in the triangular C-ring with a very slow escaping rate at room temperature or above. However, Pt nanoparticles on graphene surface keep migrating at room temperature or above which causes nanoparticle aggregation. The simulation results propose graphdiyne as an excellent substrate to stabilize Pt nanoparticle catalysts. Then, to investigate the catalytic ability of Pt nanoparticles on graphdiyne, the interaction between $O_2$ molecule and Pt nanoparticles adsorbed on graphdiyne was studied. The result indicated a potential catalytic ability of graphdiyne-Pt-nanoparticle system to the oxygen reduction reaction (ORR) in fuel cells.

**II. Simulation method**

To investigate the structures of graphene-Pt-cluster and graphdiyne-Pt-cluster systems, density functional theory (DFT) calculations were performed using the SIESTA code [19]. The norm-conserving pseudopotentials were generated using the improved Troullier-Martins scheme [20]. The generalized gradient approximation according to Perdew-Burke-Ernzerhof [21] was employed for both the generation of the pseudopotentials and the exchange-correlation functional. Non-linear exchange-correlation core corrections were used for C and Pt. The grid mesh cutoff was set 200 Ry. By above pseudopotential and double-$\zeta$ plus polarization basis set, the calculated lattice constant and cohesive energy of fcc Pt bulk was 3.97 Å and 5.76 eV, respectively, which is close to the experimental values 3.92 Å and 5.84 eV. The graphene-Pt-cluster and graphdiyne-Pt-cluster systems were simulated by a repeated slab model. The replicas of graphene or graphdiyne layers were separated by a vacuum layer of 60 Å. On the dimension parallel to the graphene or graphdiyne layer, periodical boundary was used on the edge of graphene layer. The surface area of graphene or graphdiyne layer is large enough to avoid the interaction between the replicas of Pt cluster. $1 \times 1 \times 1$ $k$-point sampling was used.

For geometry optimization and *ab initio* MD simulations, the double-ζ plus polarization basis set was applied for C atoms and the single-ζ plus polarization basis set was applied for Pt atoms to save computation time. The structures were relaxed until the Hellmann-Feynman forces are less than 0.01 eV/Å. MD simulations were performed using Nosé thermostat and Verlet algorithm with a time step of 1 fs. To find the most stable adsorption configuration of Pt cluster, i.e. the configuration with the lowest total energy, the Pt cluster was initially placed on different positions on graphene or graphdiyne surface with different molecular orientations, and geometry optimizations were performed for each structure. The adsorption energy $E_{ad}$ is defined as

$$E_{ad}=E_g+E_{Pt\text{-}cluster}-E_{g\text{-}Pt\text{-}cluster}, \qquad (1)$$

where $E_g$, $E_{Pt\text{-}cluster}$ and $E_{g\text{-}Pt\text{-}cluster}$ denote the energy of graphene (or graphdiyne), the isolated Pt cluster and the graphene-Pt-cluster (or graphdiyne-Pt-cluster) systems, respectively. To reduce the basis set superposition error in total energy calculations, for the optimized configurations the double-ζ plus polarization basis set was applied for both C and Pt atoms. To investigate this point, we calculated the adsorption energy $E_{ad}$ of Pt atom on the bridge site of graphene using the double-ζ plus polarization basis set with/without ghost orbitals, and obtained similar values $E_{ad}$=2.70/2.72 eV, respectively. This result indicates the accuracy of the double-ζ plus polarization basis set.

To obtain more information on Pt clusters adsorbed on graphene or graphdiyne surface, the ReaxFF potential [15, 16] for C-Pt systems [17, 18] was employed. The ReaxFF parameters used for C-C, Pt-Pt and C-Pt interaction were adopted from Ref. [18] besides some parameters adjusted to make the optimized structures of graphene-Pt-cluster or graphdiyne-Pt-cluster systems in consistent with DFT results. For the ReaxFF simulations on large Pt clusters, a quasi-dynamical annealing method named time-going-backward (TGB) method [22, 23], which is based on MD and has been proposed to have high annealing efficiency, was employed to find probable isomers. Then, MMC simulations were performed to find stable adsorption configurations of the probable Pt cluster isomer on graphene or graphdiyne. Finally, nudged elastic band (NEB) method [24-26] was used to find the MEPs and potential energy barriers and the rates of corresponding thermal evolutions can be estimated by the Arrhenius law.

## III. Results and discussion

### 3.1 $Pt_5$ on graphene

Initially, the adsorption of $Pt_5$ cluster on graphene was first investigated. For a $Pt_N$ cluster, the binding energy per Pt atom is defined as

$$E_b = E_{Pt\text{-atom}} - E_{Pt\text{-cluster}}/N, \tag{2}$$

where $E_{Pt\text{-atom}}$ the energy of a single Pt atom. By *ab initio* calculation, 5 Pt atoms were put closely in different positions and geometry optimization was performed. The structure of $Pt_5$ with the lowest energy is shown in the upper panel of Fig. 1(a). The calculated binding energy $E_b=3.37$ eV/atom is close to the value 3.34 eV in Ref. [27]. To investigate the most stable adsorption configuration, the simulation system was set up by putting a $Pt_5$ on a graphene sheet with periodical boundary condition. The obtained most stable adsorption configuration of $Pt_5$ is shown in the lower panel of Fig. 1(a), with the periodical boundary shown by dashed lines. According to the result, the $Pt_5$ cluster is strongly distorted with four Pt atoms bonding with the graphene sheet. The calculated $E_{ad}=4.10$ eV indicates that the adsorption of graphene to the $Pt_5$ cluster is very strong. To further investigate the thermal stability, *ab initio* MD simulation was performed at 300, 500 and 600 K. Some snapshots are shown in the upper panel of Fig. 1(b). During the simulation, the fluctuation of total energy with time kept in a certain range (the lower panel of Fig. 1(b)). In the duration of 6 ps, the $Pt_5$ cluster is always adsorbed and keeps migrating on the graphene surface, which indicates that the energy barrier of translational motion is small. During the simulation, no isomerization of $Pt_5$ occurs. The above results suggest that the adsorption configuration is very stable at room temperature. Since $E_{ad}$ is large enough, the adsorbed $Pt_5$ clusters cannot leave the graphene surface by thermal motion. However, neighboring Pt clusters may continuously hit together in their random migration on the graphene surface, causing the aggregation of small Pt clusters into large particles and loss of surface area and chemical activity.

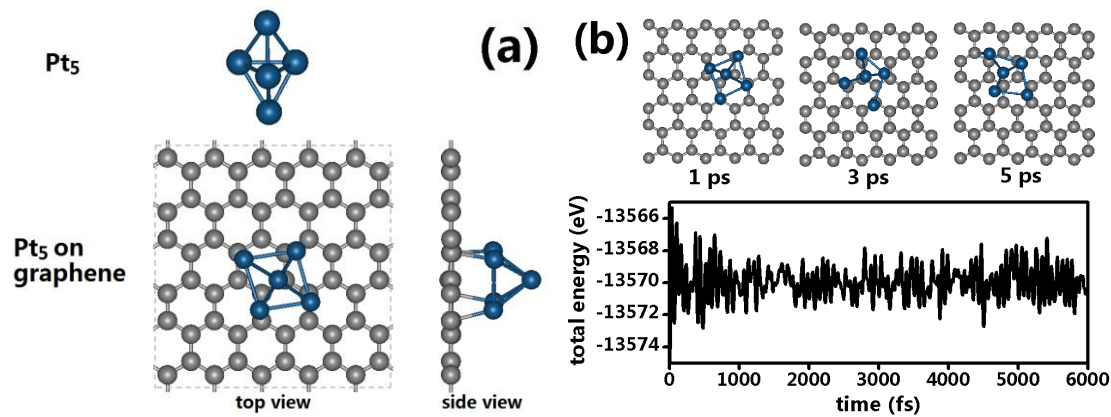

Fig. 1 (a) Structure of $Pt_5$ cluster with the lowest energy (upper panel) and the most stable adsorption configuration $Pt_5$ on graphene (lower panel). The periodical boundary is shown by dashed lines. (b) Snapshots (upper panel) and change of total energy with time (lower panel) in MD simulation at 600 K. C atom in gray and Pt in blue.

3.2 *$Pt_{13}$ on graphene*

The clusters of face centered cubic lattices, e.g. Pt or Ni cluster, have tendency to form MacKay icosahedrons with atomic magic number of 13, 55, 147 ... As the simplest MacKay icosahedrons, $Pt_{13}$ (the upper panel of Fig. 2(a)) was chosen as a typical cluster to make comparison between the adsorption ability of graphene and graphdiyne. The calculated binding energy $E_b$=4.31 eV of $Pt_{13}$ icosahedron (which is larger than that of $Pt_5$) is close to the value 4.22 eV in Ref. [27]. To investigate the stable adsorption configuration on graphene, the $Pt_{13}$ icosahedron was put close to the graphene surface in different positions and geometry optimizations were performed. The simulation system was set up as shown in the lower panel of Fig. 2(a) with periodical boundary condition shown by dashed lines. According to the result, in the adsorption configuration (the lower panel of Fig. 2(a)) obvious C-Pt bonding can be found. The adsorption energy $E_{ad}$=4.19 eV, which is close to the one of $Pt_5$, indicates that the adsorption of graphene to $Pt_{13}$ cluster is very strong.

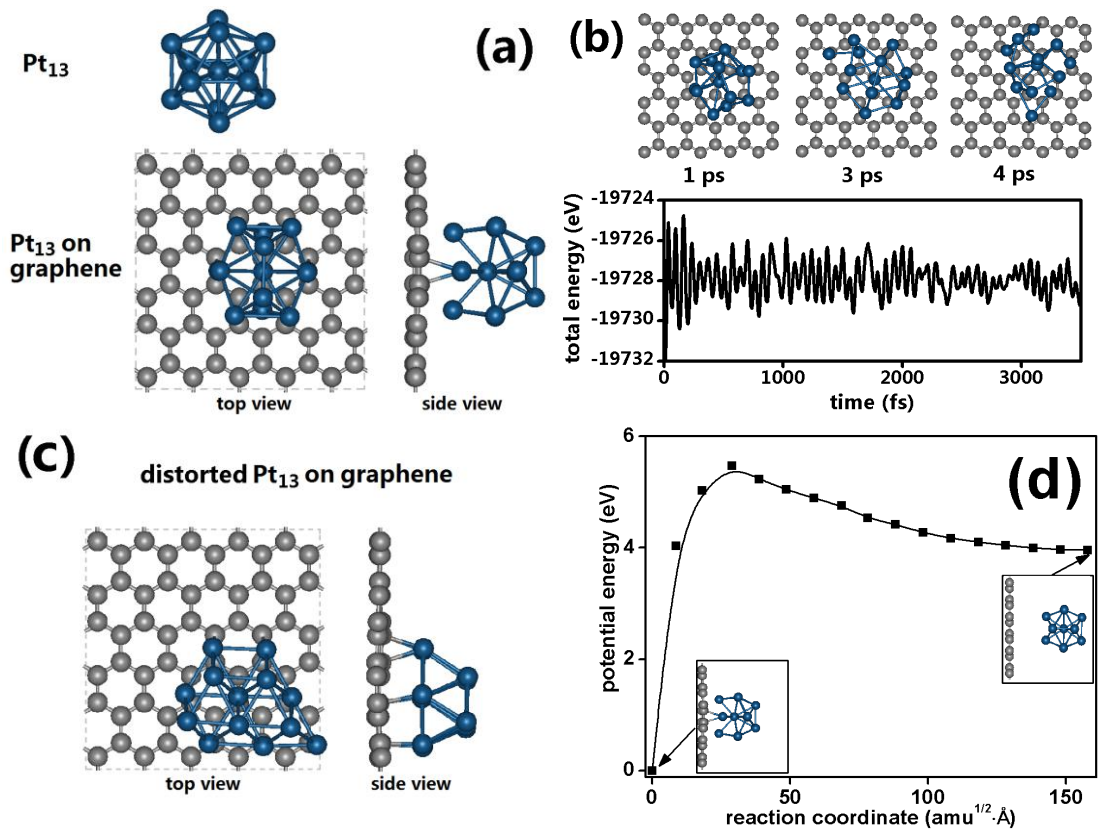

Fig. 2 (a) Structure of $Pt_{13}$ cluster with the lowest energy (upper panel) and the most stable adsorption configuration of $Pt_{13}$ on graphene (lower panel). The periodical boundary is shown by dashed lines. (b) Snapshots (upper panel) and change of total energy with time (lower panel) in MD simulation at 500 K. (c) Structure of distorted $Pt_{13}$ cluster on graphene. (d) Potential energy profile along the MEP of $Pt_{13}$ desorbing from graphene. C atom in gray and Pt in blue.

To investigate the thermal stability of $Pt_{13}$ on graphene, *ab initio* MD simulations were performed at 300, 500 and 600 K. At 300 K, the $Pt_{13}$ cluster keeps migrating on the graphene surface. At 500 K and 600 K, the $Pt_{13}$ cluster gradually distorted and transformed into a bread-like shape in a few ps. An example at 500 K is shown in the upper panel of Fig. 2(b). During the simulation, the fluctuation of total energy with time kept in a certain range (the lower panel of Fig. 2(b)), and the $Pt_{13}$ cluster was distorted at the simulation time of 4 ps. By optimizing the distorted $Pt_{13}$ cluster on the graphene surface (Fig. 2(c)), its adsorption energy was found to be $E_{ad}$=5.62 eV, which is larger than the pristine $Pt_{13}$ icosahedron. Since such transformation takes a very short time and $E_{ad}$ of distorted configuration is larger than that of original one, the distorted configuration should be more favorable for the graphene-$Pt_{13}$ system above room temperature. It is worth noting that in thermal motion the migration on the graphene, which has been found for $Pt_5$, also happens for the distorted $Pt_{13}$ cluster. By track the displacement of $Pt_{13}$ mass center relative to the graphene mass center, the average velocity of distorted $Pt_{13}$ cluster migrating on graphene was found to be about 230 m/s at 500 K. As was previously discussed in Section 3.1, such migration may be a reason for the failure of Pt-cluster-based catalyst.

To obtain more information about the graphene-$Pt_{13}$ system and save computation time, the ReaxFF potential for C-Pt systems [17, 18] was applied with some parameters adjusted for the graphene-$Pt_{13}$ system to obtain the same results as the above DFT calculations. In Ref. [18], the ReaxFF parameters were optimized against the binding structures and adsorption energies obtained from DFT calculations on the chemisorption of hydrocarbons on Pt (111). According to our calculations, the reported parameters in Ref. [18] well represent the binding energy of Pt crystals. However, using these parameters the represented binding energies $E_b$=4.14 and 5.45 eV for $Pt_5$ and $Pt_{13}$, respectively, deviate from previous *ab initio* calculations (3.37 and 4.31 eV, respectively). Therefore, we re-optimized the parameters against the DFT results of $Pt_{13}$ and the graphene-$Pt_{13}$ system. According to the DFT calculation for the icosahedronic $Pt_{13}$ cluster, the bond energy $D_e^\sigma$ and balance bong length $r_0^\sigma$ of Pt-Pt σ-bonds (for their physical meaning, see Ref. [17, 18]) were adjusted to 125 kcal/mol and 1.740 Å (original values are 98.99 kcal/mol and 1.932 Å), respectively. And according to the DFT calculation for the graphene-$Pt_{13}$ system, the bond energy $D_e^\sigma$ of C-Pt σ-bonds was adjusted to 120 kcal/mol (the original value is 141.0 kcal/mol). The adjusted parameters represented a binding energy $E_b$=4.11 eV of $Pt_{13}$ and its adsorption energy $E_{ad}$=3.96 eV on graphene, which are close to the values from DFT calculations.

To examine the validity of the re-optimized ReaxFF parameters, constant-temperature MD simulations were performed by the velocity Verlet algorithm and a time step of 0.2 fs. Simulations were initialized at a temperature $T$ and the thermal bath by Riley et al. [1] was applied, which randomly chooses an atom $i$ and replaces its velocity vector $\mathbf{V}_i^{old}$ with $\mathbf{V}_i^{new}$ in a time interval. Here,

$$\mathbf{V}_i^{new} = (1-\theta)^{1/2}\mathbf{V}_i^{old} + \theta^{1/2}\mathbf{V}^{Max}(T), \tag{3}$$

where $\mathbf{V}^{Max}(T)$ is a random velocity chosen from the Maxwellian distribution and $\theta=0.1$ is a parameter controlling the strength of velocity reset. The simulations were performed for 10 ps. At $T=300$ K, the $Pt_{13}$ cluster keeps migrating on the graphene surface. At $T\geq 500$ K, the $Pt_{13}$ cluster gradually distorted and transformed into a bread-like shape, which is just like the one in *ab initio* MD simulations, in about 1 ps. By track the displacement of $Pt_{13}$ mass center relative to the graphene mass center, the average velocity of distorted $Pt_{13}$ cluster migrating on graphene was found to be about 176 m/s at 500 K, which is close to the value 230 m/s in above *ab initio* MD simulation. The above result which is similar to that by *ab initio* MD simulations indicates the validity of the re-optimized ReaxFF parameters.

By the re-optimized ReaxFF parameters, NEB method was employed to explore the MEP of $Pt_{13}$ cluster desorbing from the graphene surface. According to the result (Fig. 2(d)), a desorption barrier $E_0 \approx 5.5$ eV was found along the MEP. Then, the rate of $Pt_{13}$ escaping from the graphene surface at temperature $T$ can be approximately evaluated by the Arrhenius law $R=R_0 exp(-E_0/kT)$, where $R_0$ is an empirical factor corresponding to atomic vibration frequency (generally in a magnitude of $10^{13}$ s$^{-1}$). At $T=300$ K, the escaping rate $R\approx 4\times 10^{-80}$ s$^{-1}$, corresponding to an average time of $\tau=1/R\approx 8\times 10^{71}$ years for $Pt_{13}$ staying on the graphene surface. At $T=1000$ K, the escaping rate $R\approx 2\times 10^{-15}$ s$^{-1}$ and the staying time of $Pt_{13}$ is even $\tau\approx 2\times 10^7$ years. Such result indicates that $Pt_{13}$ clusters adsorbed on graphene have very high thermal stability.

3.3 *$Pt_{13}$ on graphdiyne*

Graphdiyne is a newly synthesized two-dimensional layered material. In the following text, a single-layer of graphdiyne is taken as a simple model for simulations. The top-left panel of Fig. 3(a) presents the optimized structure of single-layer graphdiyne in DFT calculation, with the unit cell shown in solid line. The optimized lattice constant was found to be $a_0$=9.50 Å, in good agreement with the previous value of 9.48 Å [12] calculated by the projector-augmented-wave method. In graphdiyne, the bonds in the hexagonal rings are all $sp^2$ hybridized as in graphite, and the bonds between two hexagonal rings are –C≡C– linkages. To investigate the stable adsorption configuration on graphdiyne, the icosahedronic $Pt_{13}$ cluster was put close to the graphdiyne surface in different positions and geometry optimizations were performed. The simulation system was set up as shown in the lower panel of Fig. 3(a) with periodical boundary condition shown by dashed lines. According to the result, in the adsorption configuration (the lower panel of Fig. 3(a)) the $Pt_{13}$ cluster locates in the large C ring of graphdiyne and obvious C-Pt bonding can be found. The adsorption energy $E_{ad}$=8.93 eV, which is much larger than that of graphene surface, indicates that the bonding between graphdiyne and $Pt_{13}$ cluster is very strong. Such strong adsorption ability of graphdiyne should be attributed to the additional $p_x$-$p_y$ π/π* states existing in the –C≡C– bonds. This enables the π/π* states to rotate towards any direction perpendicular to the line of –C≡C–, thus making it possible for the π/π* states from the –C≡C– bonds at a given acetylenic ring to all point towards the $Pt_{13}$ cluster. To present the detail, the partial density of states (PDOS) of $p_x$+$p_y$ in the –C≡C– bonds of graphdiyne and the PDOS of surface atoms of $Pt_{13}$ are plotted in Fig. 3(e). The states above the Fermi energy of graphdiyne should be the π* states, which just locate below the Fermi energy of $Pt_{13}$. So, when graphdiyne combines with $Pt_{13}$, the charge transfers from Pt cluster to the π-orbitals on the large C ring. To prove this, Hirshfeld charge analysis was performed for the graphdiyne-$Pt_{13}$ and graphene-$Pt_{13}$ systems. In the graphdiyne-$Pt_{13}$ system, the calculated charge transfer 2.08 $e$ from $Pt_{13}$ to graphdiyne, which is larger than the value 1.34 $e$ of graphene-$Pt_{13}$ system, verifies the above proposition.

To investigate the thermal stability of $Pt_{13}$ on graphdiyne, *ab initio* MD simulations were performed at 300, 500 and 600 K. Some snapshots at 500 K are shown in the upper panel of Fig. 3(b). During the simulation, the fluctuation of total energy with time kept in a certain range (the lower panel of Fig. 3(b)). At every temperature, the $Pt_{13}$ cluster keeps moving around the equilibrium position and never leaves the large C ring of graphdiyne. The above simulation result further indicates the strong bonding between graphdiyne and $Pt_{13}$ cluster. $Pt_{13}$ clusters should be very stable on graphdiyne surface without thermal migration at room temperature.

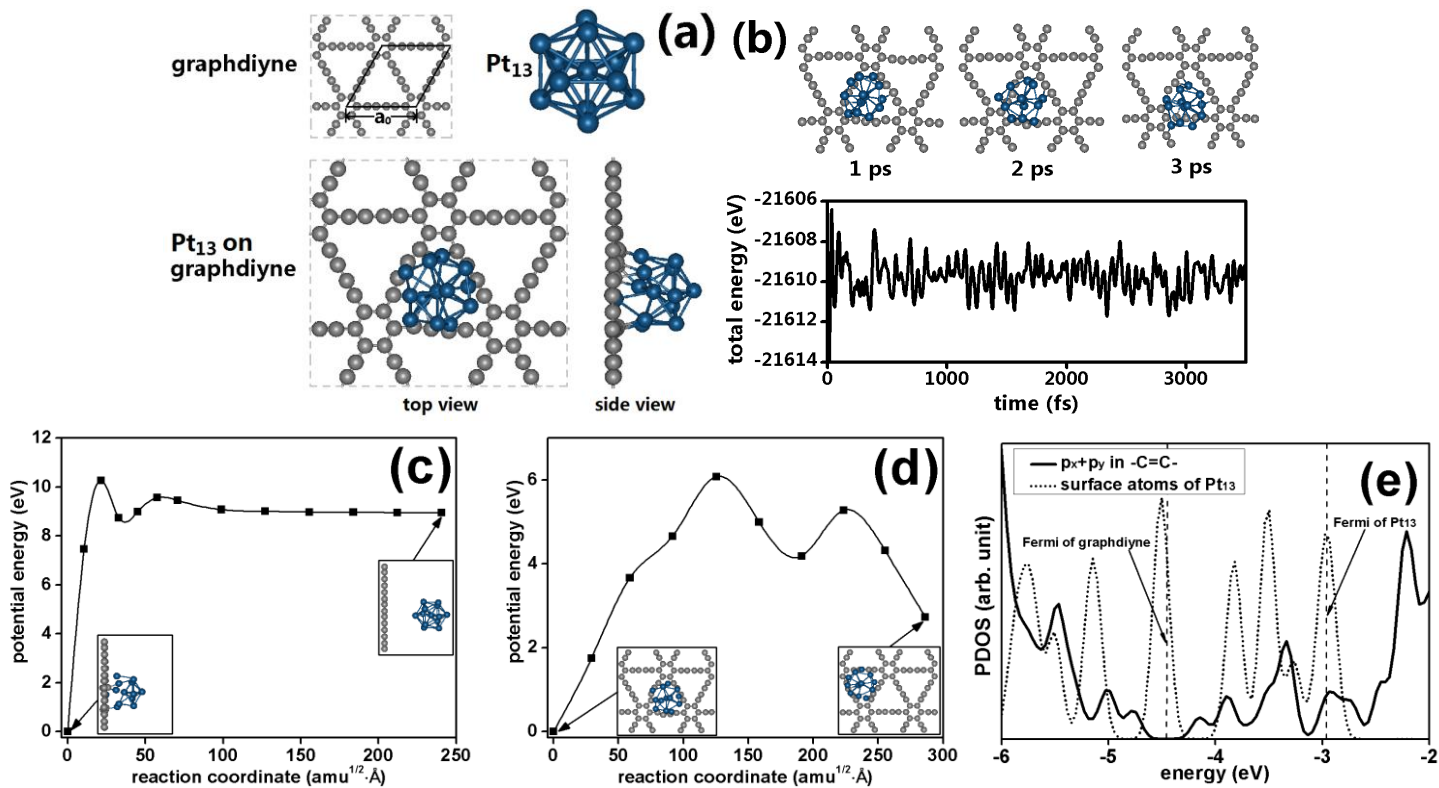

Fig. 3 (a) Structure of graphdiyne (top-left panel), the structure of $Pt_{13}$ cluster with the lowest energy (top-right panel) and the most stable adsorption configuration of $Pt_{13}$ on graphdiyne (lower panel). The periodical boundary is shown by dashed lines. (b) Snapshots (upper panel) and change of total energy with time (lower panel) in MD simulation at 500 K. (c) Potential energy profile along the MEP of $Pt_{13}$ desorbing from graphdiyne. (d) Potential energy profile along the MEP of $Pt_{13}$ migrating in the graphdiyne surface. C atom in gray and Pt in blue. (e) The PDOS of $p_x+p_y$ in the –C≡C– bonds of graphdiyne (solid line) and the PDOS of surface atoms of $Pt_{13}$ (dotted line). The vertical dashed lines present the Fermi energy of graphdiyne and $Pt_{13}$.

Using the ReaxFF potential, the processes of $Pt_{13}$ desorbing from the graphdiyne and migrating in the graphdiyne surface were studied. According to the re-optimization procedure similar to that for the graphene-$Pt_{13}$ system, the bond energy $D_e^\sigma$ of C-Pt σ-bonds in the graphdiyne-$Pt_{13}$ system was adjusted to 140 kcal/mol (the original value is 141.0 kcal/mol), and the bond energy $D_e^\sigma$ and balance bond length $r_0^\sigma$ of Pt-Pt σ-bonds in the icosahedronic Pt$_{13}$ cluster were the same as previous used values (125 kcal/mol and 1.740 Å, respectively). The re-optimized parameters represented an adsorption energy $E_{ad}$=8.96 eV of Pt$_{13}$ on graphdiyne, which are close to the value from DFT calculations. By the NEB method, the MEP of desorption process is plotted in Fig. 3(c), showing a barrier $E_0$≈10.3 eV which is much higher than that of Pt$_{13}$ on graphene. By the Arrhenius law $R=R_0 exp(-E_0/kT)$ ($R_0$~$10^{13}$ s$^{-1}$), the rate $R$ of Pt$_{13}$ escaping from the graphdiyne surface at temperature $T$ can be approximately evaluated. At $T$=300 K, the escaping rate $R$≈1×10$^{-160}$ s$^{-1}$, corresponding to an average time of $\tau=1/R$≈3×10$^{152}$ years for Pt$_{13}$ staying on the graphdiyne surface. At $T$=1000 K, the escaping rate $R$≈1×10$^{-39}$ s$^{-1}$ and the staying time of Pt$_{13}$ is even $\tau$≈3×10$^{31}$ years. For the migration of Pt$_{13}$ on graphdiyne, at $T$=300 ~ 1000 K the barrier $E_0$≈6.1 eV corresponds to a rate $R$≈3×10$^{-90}$ ~ 2×10$^{-18}$ s$^{-1}$, corresponding to an average time of $\tau=1/R$≈1×10$^{82}$ ~ 2×10$^{10}$ years. To verify the above result, MD simulations were performed using the ReaxFF potential with the method in Section 3.2 (velocity Verlet algorithm and a time step of 0.2 fs). The simulations were performed for 10 ps at 500, 600 and 800 K. During the simulations, the Pt$_{13}$ cluster never migrates or escapes, only keeps moving around the equilibrium position on graphdiyne. The above results indicate that Pt$_{13}$ clusters adsorbed on graphdiyne have very high thermal stability. The adsorbed Pt$_{13}$ clusters can neither escape from graphdiyne nor migrate in the graphdiyne surface.

### 3.4 Pt$_{48}$ on graphene and graphdiyne

In practical applications, Pt nanoparticles used as catalyst usually have a size of 2~5 nm. Therefore, we chose Pt$_{48}$ as a larger and more actual model. To obtain the stable configuration of Pt$_{48}$ nanoparticle, a systematic procedure based on the TGB annealing method was applied using the ReaxFF potential. The theoretical background and simulation technique can be seen in Ref. [22, 23], and Fig. 4(a) illustrates the detail of the procedure. Starting from a cubic Pt$_{48}$ crystal (denoted as **I$_0$**), 8 independent runs of the TGB method were performed to anneal the system. After the annealing, 8 different $Pt_{48}$ isomers were obtained and the one with the lowest potential energy was denoted as $I_1$. In the next round, 8 new runs of the TGB method were performed starting from $I_1$, and in the products the isomer with the lowest potential energy was denoted as $I_2$. Such process was carried out repeatedly. After 7 rounds, no isomers with lower potential energy, i.e. higher binding energy $E_b$, than the last round were found (the left panel of Fig. 4(a)). So, the isomer $I_7$ should get close to actual $Pt_{48}$ nanoparticle in catalyst.

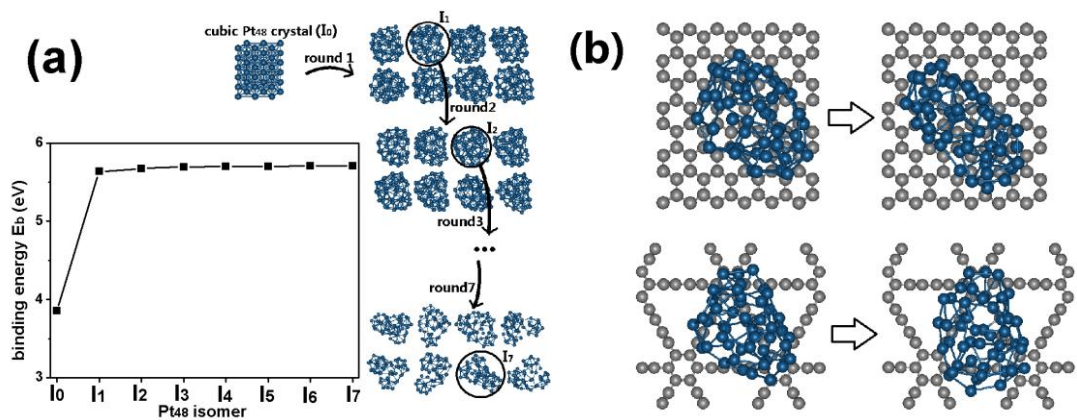

Fig. 4 (a) Sketch for the annealing procedure of $Pt_{48}$ starting from cubic $Pt_{48}$ crystal (denoted as $I_0$). In every round of annealing, 8 independent runs of the TGB method were performed. $I_1 \sim I_7$ denote the isomer with the lowest potential energy after each round (marked by circles). The left panel plots the binding energy $E_b$ of $I_0 \sim I_7$. (b) Sketch for MMC simulation of the isomer $I_7$ on graphene (upper panel) and graphdiyne (lower panel). C atom in gray and Pt in blue.

To find the configuration of $Pt_{48}$ absorbed on graphene, the ReaxFF potential was employed using the parameters adjusted for graphene-$Pt_{13}$ system. First, geometry optimization was performed for the isomer $I_7$ putting on the graphene surface with a random orientation. Then, MMC simulation was performed for the optimized configuration at 900 K, i.e. close to the melting point of small Pt nanoparticles, to let the isomer relax on the graphene surface (the upper panel of Fig. 4(b)). Finally, geometry optimization was performed again and we got a more stable adsorption configuration than the original one. To find the most stable adsorption configuration, the above procedure was performed repeatedly, starting from different orientations of the isomer $I_7$. According to the result, $E_{ad}$=8.7 eV for $Pt_{48}$ on graphene indicates that the adsorption of graphene to large Pt nanoparticle is strong.

To investigate the adsorption of Pt nanoparticles on graphdiyne, a simulation procedure similar to that for graphene was applied on the graphdiyne-$Pt_{48}$ system. The adsorption of the isomer $I_7$ on graphdiyne with different orientations was explored. The MMC step is shown in the lower panel of Fig. 4(b). According to the result, the most stable configuration of $Pt_{48}$ on graphdiyne was found to have an adsorption energy of $E_{ad}$=13.1 eV, which is much larger than that for graphene. This result indicates that graphdiyne is much better than graphene to stabilize Pt nanoparticles.

3.5 *$O_2$ on Pt catalyst adsorbed on graphdiyne*

According to above results, Pt nanoparticles can be stabilized on graphdiyne surface and the aggregation and loss of catalytic ability can be avoided. Then, the catalytic ability of Pt nanoparticles on graphdiyne should be investigated. In fuel cell, the rate of ORR at the cathode is one of the limits for better performance [28, 29]. Therefore, in this section the adsorption of $O_2$ molecule was taken as a typical example to investigate the catalytic ability of Pt nanoparticles on graphdiyne.

The simulation system was built by putting an $O_2$ molecule on the graphiyne-$Pt_{13}$ system. To find most stable adsorption configurations, geometry optimizations were performed for the $O_2$ molecule put on the $Pt_{13}$ cluster in different positions and orientations. The upper panel of Fig. 5(a) shows four different adsorption configurations with lowest potential energy, which are marked as **1**, **2**, **3** and **4**. The adsorption energies $E_{ad}=E_{g-Pt13}+E_{O2}-E_{g-Pt13-O2}$ in the position **1, 2, 3** and **4** are 1.38, 2.09, 1.47 and 1.51 eV, respectively. And the O-O bond lengths in the position **1, 2, 3** and **4** are 1.291, 1.354, 1.286 and 1.295 Å, respectively. The fact that the bond lengths are all longer than the value in the gas phase (1.233 Å) indicates the weakening of O-O bond by the O-Pt bonding. According to the result, the position **2** is the most stable with the highest adsorption energy and longest O-O bond length.

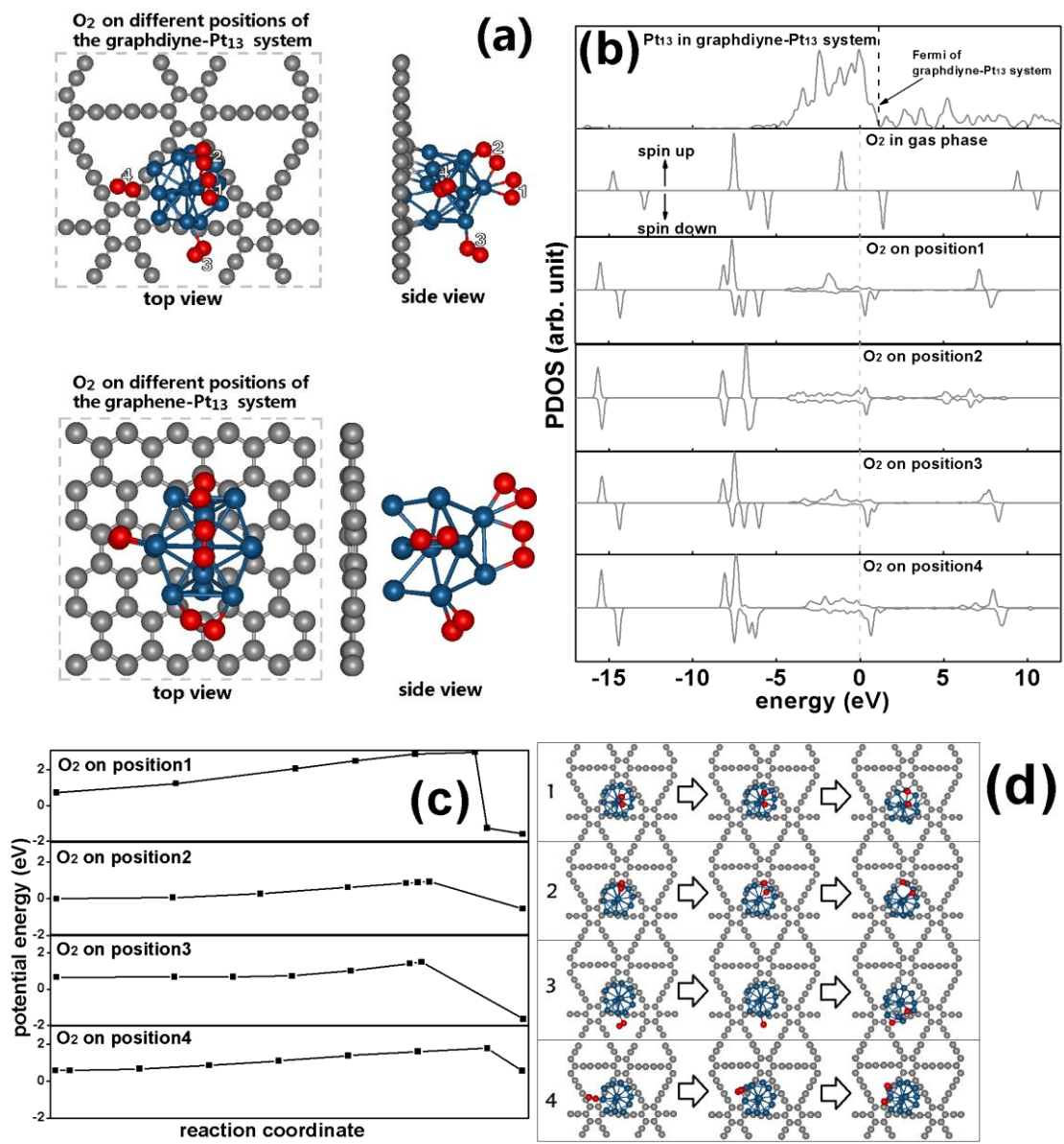

Fig. 5 (a) Upper panel: most stable adsorption configurations of $O_2$ on graphdiyne-$Pt_{13}$ system. **1**, **2**, **3** and **4** indicate four different configurations. Lower panel: most stable adsorption configurations of $O_2$ on graphene-$Pt_{13}$ system. The periodical boundary is shown by dashed lines. C atom in gray, Pt in blue and O in red. (b) Spin-polarized PDOS for $Pt_{13}$ in the graphdiyne-$Pt_{13}$ system, $O_2$ in the gas phase and $O_2$ adsorbed in the positions **1**, **2**, **3** and **4** of the graphdiyne-$Pt_{13}$ system. For $O_2$ in the gas phase and $O_2$ adsorbed on the graphdiyne-$Pt_{13}$ system, the Fermi energy is set to zero. For the graphdiyne-$Pt_{13}$ system, the energy zero is set equal to the Fermi energy of $O_2$ in the gas phase. (c) The potential energy profiles of $O_2 \rightarrow 2O$ along reaction paths starting from the configurations **1**, **2**, **3** and **4**. The potential energy of **2** is set to zero. (d) The reactants, transition states and products of $O_2 \rightarrow 2O$, starting from the configurations **1**, **2**, **3** and **4**, respectively.

To further understand the bonding of $O_2$ on the graphdiyne-$Pt_{13}$ system, we plotted the PDOS of $O_2$ molecule and $Pt_{13}$ in the graphdiyne-$Pt_{13}$ system in Fig. 5(b).

For an $O_2$ molecule in the gas phase, the HOMO and LUMO are spin-up and spin-down antibonding $2\pi^*$ orbitals (see the 2$^{nd}$ panel of in Fig. 5(b)), respectively. By contrast, the Fermi energy of graphdiyne-$Pt_{13}$ system is higher (see the 1$^{st}$ panel of in Fig. 5(b)), getting close to the LUMO of $O_2$ in the gas phase. So, when $O_2$ combines with $Pt_{13}$, the electrons in $Pt_{13}$ should partially flow into $O_2$, leading to a decrease of the total energy and an enlarged O–O bond length. By Hirshfeld charge analysis, the charge transfer from $Pt_{13}$ to $O_2$ was calculated to be 0.23, 0.42, 0.24 and 0.30 *e* in the position **1**, **2**, **3** and **4**, respectively. Such charge transfer leads to a lift of Fermi energy. In the 3$^{rd}$~6$^{th}$ panels of Fig. 5(b), it can be seen that the Fermi energies of graphdiyne-$Pt_{13}$–$O_2$ systems get closer to the LUMO of $O_2$, i.e. more electrons populate into $O_2$. Comparing the charge transfer to the O-O bond length, we can see the trend that the more electrons that are populated to the LUMO of $O_2$, the more the adsorption energy increases. The position **2**, which has the largest charge transfer, is the most stable configuration with longest O-O bond length. The weakening of O-O bond indicates the potential catalytic ability of Pt nanoparticles on graphdiyne.

Then, to investigate the detail about ORR on graphdiyne-$Pt_{13}$ system, we focused on the first step of the reaction $O_2 \rightarrow 2O$. The potential energy profiles of $O_2 \rightarrow 2O$ along reaction paths (Fig. 5(c)) were plotted by geometry optimization with increasing O-O bond length, starting from the configurations **1**, **2**, **3** and **4** (the potential energy of **2** is set to zero). The corresponding reactants, transition states and products of $O_2 \rightarrow 2O$ are shown in Fig. 5(d). According to the result, the reaction barriers for $O_2 \rightarrow 2O$ from **1**, **2**, **3** and **4** are 2.27, 0.93, 0.85 and 1.20 eV, respectively. For configuration **1** and **4**, their charge transfers from $Pt_{13}$ to $O_2$ is smaller than **2**, and their reaction barriers are higher than **2**. However, the charge transfer for configuration **3** is also smaller than **2**, but its reaction barrier is even a bit lower than **2**. At room temperature, the ORR reaction via **2** or **3** should be probable.

Finally, the adsorption of $O_2$ molecule on graphdiyne-$Pt_{13}$ system was compared with the graphene-$Pt_{13}$ system which can be regarded as a typical model of common used C-Pt catalysts. So, geometry optimizations were performed for the $O_2$ molecule put on the graphene-$Pt_{13}$ system in different positions and orientations (lower panel of Fig. 5(a)). According to the result, the fact that the O-O bond lengths in the graphene-$Pt_{13}$ system (in the range of 1.299~1.367 Å) are close to that in graphdiyne-$Pt_{13}$ system (1.291~1.354 Å) indicates the potential catalytic ability of graphdiyne-$Pt_{13}$ system. In graphene-$Pt_{13}$ system, the charge transfers from $Pt_{13}$ to $O_2$ are in the range of 0.27~0.44 *e*, which is also close to that in graphdiyne-$Pt_{13}$ system. This fact also indicates the catalytic ability of graphdiyne-$Pt_{13}$ system.

## IV. Conclusion

In this work, the adsorption and thermal migration of Pt nanoparticles on graphdiyne and graphene substrate were theoretically investigated. Using multiscale calculations by *ab initio* method and the ReaxFF potential, geometry optimizations, MD simulations, MMC simulations and MEP calculations were performed to investigate the adsorption energy, the desorption and migration rate of Pt nanoparticles on graphdiyne and graphene. For graphene substrate, the adsorption to Pt nanoparticles is strong enough. At room temperature or above, Pt nanoparticles can stably stay on graphene surface with a very slow escaping rate, but keep migrating and hitting together, resulting in aggregating into large nanoparticles and loss of surface area and chemical activity. For graphdiyne substrate, the adsorption to Pt nanoparticles is even stronger than graphene due to the charge transfer from Pt to the π-orbitals of the large C ring. Therefore, Pt nanoparticles on graphdiyne are more stable than on graphene at room temperature or above. Furthermore, the porous structure of graphdiyne can avoid the thermal migration of Pt nanoparticles. According to the simulation results, Pt nanoparticles are stabilized on graphdiyne substrate without aggregation and loss of chemical activity. Finally, the interaction between $O_2$ molecule and $Pt_{13}$ cluster adsorbed on graphdiyne was investigated. Since electrons in $Pt_{13}$ populate into the LUMO of $O_2$, the O-O bond is weakened. The result indicates that Pt nanoparticles on graphdiyne have a potential catalytic ability to the ORR in fuel cells. According to above results, graphdiyne is proposed to be an excellent substrate for Pt nanoparticle catalyst to replace graphite or amorphous carbon matrix.

***

**Acknowledgements**

This work was supported by the National Natural Science Foundation of China under Grant No. 11304239, and the Fundamental Research Funds for the Central Universities.